\documentclass[twocolumn,prl,showpacs,aps]{revtex4-1}
\usepackage{amssymb}
\usepackage{epsfig,amsmath}
\usepackage{graphicx}% Include figure files
\usepackage{dcolumn}% Align table columns on decimal point
\usepackage{bm}

\usepackage{environ}
\NewEnviron{Leq}{%
\begin{equation}
\scalebox{2.5}{$\BODY$}
\end{equation}}

\newcommand{\beq}{\begin{equation}}
\newcommand{\eeq}{\end{equation}}
\newcommand{\beqa}{\begin{eqnarray}}
\newcommand{\eeqa}{\end{eqnarray}}

\newcommand{\ra}{\rangle}

\def\ajp#1{{ Am.\ J.\ Phys.} {\bf #1}}

\def\jmo#1{{ J.\ Mod.\ Opt.} {\bf#1}}
\def\jpb#1{{ J.\ Phys.\ B} {\bf#1}}

\def\njp#1{{ New\ J.\ Phys.} {\bf#1}}

\def\pra#1{{ Phys.\ Rev. A\/} {\bf#1}}

\def\prl#1{{ Phys.\ Rev.\ Lett.} {\bf#1}}

\def\sci#1{{ Science} {\bf#1}}
\def\rmp#1{{ Rev. \ Mod. \ Phys.} {\bf#1}}

\begin{document}

 %\linenumbers
 %\lipsum

\title{Entanglement Polygon Inequality in Qubit Systems}
\author{Xiao-Feng Qian$^{1,2}$}
\email{xiaofeng.qian@rochester.edu}
\author{Miguel A. Alonso$^{1,2,4}$}
\author{J.H. Eberly$^{1,2,3}$}
\affiliation{$^{1}$Center for Coherence and Quantum Optics, $^{2}$The Institute of Optics, $^{3}$Department of Physics \& Astronomy, University
of Rochester, Rochester, New York 14627, USA,\\ $^{4}$Aix-Marseille Univ., SNRF, Centrale Marseille, Institut Fresnel, UMR 7249, 13397 Marseille Cedex 20, France}

\date{\today }


\begin{abstract}
We prove a set of tight entanglement inequalities for arbitrary $N$-qubit pure states. By focusing on all bi-partite marginal entanglements between each single qubit and its remaining partners, we show that the inequalities provide an upper bound for each marginal entanglement, while the known monogamy relation establishes the lower bound. The restrictions and sharing properties associated with the inequalities are further analyzed with a geometric polytope approach, and examples of three-qubit GHZ-class and W-class entangled states are presented to illustrate the results.
\end{abstract}

\maketitle

%\begin{Leq}
%a+b=c 
%\end{Leq}

%##############################

%\section{Introduction}
\noindent{\bf Introduction:} Entanglement is a special type of correlation among physical systems. Its restriction and its distribution as a resource among multiple parties both play an important role in the proposals of various quantum information and technology tasks \cite{Nielsen-Chuang-00}. Various entanglements (concerning different compositions of entangled parties) exist in a multiparty system, and different aspects of entanglement distribution can be considered. For example, in a three-qubit system, there exist six different bi-partite entanglements $E_{1|2}$, $E_{2|3}$, $E_{1|3}$, $E_{1|23}$, $E_{2|31}$, $E_{3|12}$. Here $E_{A|B}$ denotes bi-partite entanglement between parties $A$ and $B$, where each party can contain either one or the remaining (two) qubits. Coffman, Kundu, and Wootters initiated the focus of distribution from a ``one-to-group" entanglement (between a singled-out qubit and a group of qubits) into all ``one-to-one" entanglements (between the singled-out qubit and each individual qubit in the group) \cite{Coffman-etal-00}. This has led to the discovery of the well known entanglement monogamy relation, $E^2_{1|23} \ge E^2_{1|2}+E^2_{1|3}$, followed by various $N$-party extensions \cite{Osborne-Verstraete-06, Lohmayer-etal-06, Ou-Fan-07, Hiroshima-etal-07, Eltschka-etal-09, Giorgi-11,Streltsov-etal-12, Bai-etal-14, Regula-etal-14, Luo-etal-16}.

Monogamy relations reveal one aspect of fundamental connections among a particular set of bipartite entanglements in a multiparty system. Here we focus on a different aspect of entanglement distribution by considering a set of same-type bipartite entanglements. Such a consideration reveals a different aspect of fundamental entanglement restriction. It can be crucial to various proposals to run multiple parallel entanglement-assisted quantum tasks in a single multiparty system \cite{Nielsen-Chuang-00, Horodecki-etal-09}, for example, quantum information transfer from one site to another in a multi-site spin chain system \cite{Chen-etal}. 

Specifically, we consider the restrictions among all ``one-to-group" entanglements between a single qubit and the remaining ones in an arbitrary $N$-qubit system. Such bi-partite entanglements are sometimes also called quantum marginal entanglements, as discussed by Walter et al.~\cite{Walter-etal-13}. For example, in the three-qubit case, the concerned marginal entanglements are $E_{1|23}$, $E_{2|31}$, and $E_{3|12}$. We obtain a generic set of entanglement restriction relations that can be called polygon inequalities for arbitrary pure states in terms of generic entanglement measures including von Neumann entropy $S$ \cite{von-Neumann-32}, concurrence $C$ \cite{Wootters-98}, negativity $N$ \cite{Vidal-Werner-02}, and a normalized Schmidt weight $Y$ \cite{K}. We then show each entanglement polygon inequality provides an upper bound for a corresponding ``one-to-group" marginal entanglement, while the monogamy relation provides its lower bound. We further illustrate these inequalities with a geometric representation to give a clear visualization of the restriction and sharing properties. \\

% \noindent{\bf Entanglement Sharing Restriction} \quad
 %\section{Entanglement Polygon Inequality} 
 \noindent{\bf Entanglement Polygon Inequality:} We consider all $N$ ``one-to-group" entanglements in an $N$-qubit system, i.e., $E_{1|23...N}$, $E_{j|1...k\neq j...N}$, ..., $E_{N|12...N-1}$. From now on we use the notation $E_{j}$ to represent $E_{j|1...k\neq j...N}$ for simplicity. Here we take all these entanglements $\{E_j\}$ being normalized, i.e., varying between 0 and 1. Any connection among these entanglements has to be restricted by the underlying states, which are governed by the fundamentals of quantum mechanics. Therefore we consider an arbitrary $N$-qubit pure state that in general can be expressed as
\begin{equation}
|\Psi\rangle
=\sum_{s_{1},...,s_{N} = 0,1} c_{s_{1},...,s_{N}}|s_{1}\rangle
...|s_{N}\rangle ,  \label{pure state}
\end{equation}%
where $c_{s_{1},...,s_{N}}$ are normalized coefficients and $s_{j}$
takes the value 0 or 1 corresponding to the two states $|0\rangle $,
$|1\rangle $ of the $j$-th qubit, with $j=1,2,3,...,N$.

Our main result is a set of inequalities:
\begin{equation} \label{polygon-ineq}
E_{j} \leq  \sum_{k\neq j}E_{k},
\end{equation}
among all $N$ ``one-to-group" marginal bi-partite entanglements $E_j$. The inequality is valid for arbitrary $N$-qubit pure states as given in (\ref{pure state}). Here $E_{j}$ can be any one of many entanglement measures including von Neumann entropy \cite{von-Neumann-32}, concurrence \cite{Wootters-98}, and negativity \cite{Vidal-Werner-02}, as well as the normalized Schmidt weight \cite{K, Qian-Eberly-10, AQE-16}.

In the one-qubit case, $N=1$, the inequality reduces to $E_1\le0$, which means $E_1=0$ due to the non-negativity of entanglement measures. It is obviously true that there is no entanglement for a single qubit. In the two-qubit case, $N=2$, the inequality becomes $E_1\le E_2$ and $E_2\le E_1$, which means $E_1=E_2$. This is apparent for any two-qubit state, i.e., the entanglement between qubit one and qubit two should always be equal to the entanglement between qubit two and qubit one. 

\begin{figure}[h]
\includegraphics[width=8cm]{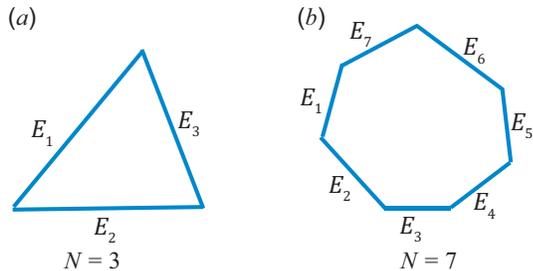}
\caption{Schematic illustration of entanglement polygon inequalities. Closed $N$-sided polygons are shown in (a) and (b) to illustrate entanglement restriction (\ref{polygon-ineq}) for $3$ and $7$ qubits respectively. The length of each side represents correspondingly the value of a marginal entanglement $E_j$.}
\label{polygon}
\end{figure}

The inequality begins to become non-trivial with increasing number of qubits $N\ge3$. To have a superficial understanding, one can assume that the value of each entanglement $E_j$ represents the length of a line. Then the above set of inequalities (\ref{polygon-ineq}) guarantees that these lines can form a closed $N$-sided polygon. See Fig.~\ref{polygon} for a schematic illustration for $N=3$ and $N=7$. Therefore one can naturally call such an entanglement restriction an entanglement polygon inequality.

To have a different understanding of the restriction relation (\ref{polygon-ineq}), one can shift to the perspective of resource sharing. Consider the distribution of a given amount of total entanglement. When one adds $E_j$ to both sides of  (\ref{polygon-ineq}) and divides by 2, one immediately obtains
\beq \label{sharing}
E_j\le E_T/2,
\eeq
where $E_T=\sum_{j=1}^N E_j$ is the total of all individual entanglements. In the point of view of entanglement as a resource, the above relation simply says that no individual participant $E_j$ gets more than half of the total. 

This is a resource sharing rule for all the participating entanglements. It limits the flexibility of distributing a given total resource. Such a sharing restriction will  be very helpful to propose appropriate multiple quantum information tasks in a single multiparty system, and to guide designs to avoid overloading tasks on any particular entanglement. A detailed understanding and analysis of the inequalities (\ref{polygon-ineq}) and  (\ref{sharing}) will be discussed in the following with a visualizable geometric representation.

As pointed out in the beginning, the well-known monogamy relation \cite{Coffman-etal-00} and the entanglement polygon inequalities (\ref{polygon-ineq}) concern different sets of bi-partite entanglements of a multiparty system. However, there is a common element in both restriction relations, i.e., the ``one-to-group" or marginal entanglement $E_j$ based on the same bi-partition. The $N$-qubit version of the monogamy relation reads
\beq \label{monogamy}
E^2_{j} \geq \sum_{k \neq j} E^2_{j|k},
\eeq
which is also valid for a generic entanglement measure $E$ such as concurrence \cite{Osborne-Verstraete-06}, negativity \cite{Ou-Fan-07} and von Neumann entropy \cite{Luo-etal-16}. 
 
By taking the square root of the monogamy relation (\ref{monogamy}), and combining with the entanglement polygon inequality (\ref{polygon-ineq}), one immediately finds the interesting relation  
\beq
\sqrt{\sum_{k \neq j} E^2_{j|k}}\le E_j\le\sum_{k\neq j} E_k.
\eeq
Obviously, the traditional monogamy relation provides a lower bound for the marginal entanglement $E_{j}$ while the entanglement polygon inequality establishes its upper bound.\\

%\section {Inequality Proof}
 \noindent{\bf Inequality Proof:} The complete proof of the entanglement polygon inequality (\ref{polygon-ineq}) for various entanglement measures is non-trivial, and its details are given in the Appendix. Here we provide a brief sketch of the proof as an illustration of the strategy. The first step is to prove that inequality (\ref{polygon-ineq}) holds for a specific entanglement monotone $Y$ \cite{Qian-Eberly-10}, i.e.,  
 \beq
 Y=1-\sqrt{\frac{2}{K}-1}.
 \eeq
It is the normalized version of Schmidt weight \cite{K} 
\beq
K=\frac{1}{\lambda_1^2+\lambda_2^2}
\eeq
defined based on the Schmidt coefficients $\sqrt{\lambda_1}, \sqrt{\lambda_2}$ of a two-party pure state (\ref{pure state}), of which one party is taken as a single qubit and the second party contains all the remaining qubits \cite{Schmidt, Fedorov-Miklin-14, Ekert-Knight-95}.

The second step is to show that different entanglement measures, i.e., von Neumann entropy $S$ \cite{von-Neumann-32}, concurrence $C$ \cite{Wootters-98}, and negativity $N$ \cite{Vidal-Werner-02}, are all concave and monotonically increasing functions of $Y$ in the region [0,1]. Then a function $E(Y)$ can be used to represent a generic entanglement measure. 
 
The third step is to combine the results of the first two steps. In this final step we assume without loss of generality that Max$\{Y_i\}=Y_j$, with $i=1,2,3,...,N$. Thus the monotonic increasing property of the function $E(Y)$ ensures the relation Max$\{E(Y_i)\}=E(Y_j)$. We also define a linear function $f(Y)=E(Y_j)Y/Y_j$, and the concavity property of $E(Y)$ guarantees 
\beq \label{concave}
f(Y_{k\ne j})\le E(Y_{k\ne j}).
\eeq

Then we prove the relation (see details in Appendix)
\beq \label{f(Y)}
\sum_k f(Y_{k\ne j})\ge E(Y_j),
\eeq 
by using the result of the first step, i.e., $Y_j\le \sum_k Y_{k\ne j}$. By combining (\ref{concave}) and (\ref{f(Y)}), it is then straightforward to see that the entanglement polygon inequality (\ref{polygon-ineq}) is valid for these generic measures.

The inequality (\ref{polygon-ineq}) in terms of $Y$ is uniquely tight, i.e., it not only applies to all $N$-qubit pure states, but additionally those states exhaust the inequality, occupying its interior and also its boundaries. The inequality (\ref{polygon-ineq}) in terms of $S$, $C$, and $N$ is looser than that in terms of Y due to the concave properties. \\

\begin{figure}[h!]
\includegraphics[width=8 cm]{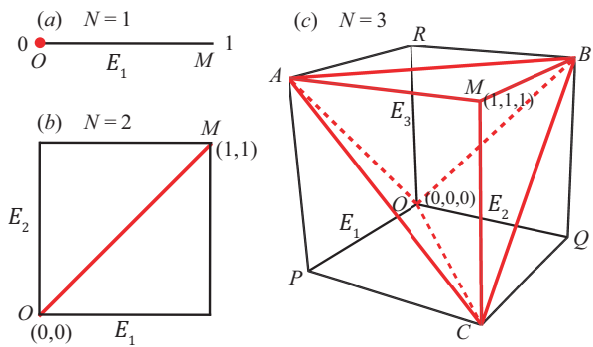}
\caption{$N$-dimensional spaces in which the point ${\bf E}$ is
defined, corresponding to (a) $N=1$ (unit line segment), (b) $N=2$
(unit square), and (c) $N=3$ (unit cube). In all cases, the point $O$
corresponds to no entanglement and the point $M$ to maximal
entanglement.}
\label{N123}
\end{figure}

%%%%%%%%%%%%%%
%\section{Polytope Analysis }  
\noindent{\bf Polytope Analysis:} We now further analyze the entanglement polygon inequality with a geometric approach that captures both the restriction (\ref{polygon-ineq}) and resource sharing  (\ref{sharing}) features. First we assume the limiting case where there is no restriction among all $N$ different entanglements $E_{j}$; they are then independent of each other and can be used to identify axes $E_{j}$ in a unit $N$-dimensional hypercube ($0\le E_{j} \le 1$). Each combination of all entanglements $\{E_j\}$ represents a unique point ${\bf E} = (E_{1},E_{2},...,E_{N})$ inside this hypercube. For example, when $N=1,2,3$ the sets of entanglements $\{E_j\}$ form a line segment, square, and cube respectively (see Fig.~\ref{N123}). 

It is natural to imagine that, under the restriction relation (\ref{polygon-ineq}), the occupied hypervolume will be reduced. This is indeed the case. For example, it will reduce to a single point $E_1=0$ and a single diagonal line $E_1=E_2$ respectively in the trivial cases of $N=1$ and $N=2$ as shown in Fig.~\ref{N123} by the red dot $O$ and red line $OM$.  

As a non-trivial illustration we analyze in detail the case of $N=3$, for which the generic entanglement polygon inequality (\ref{polygon-ineq}) implies
\begin{equation}
E_{1}+E_{2}\geq E_{3}, \quad
E_{2}+E_{3}\geq E_{1}, \quad
E_{3}+E_{1}\geq E_{2}.\label{3inequality}
\end{equation}
One notes that when the three inequalities all take the equal sign, each of them
defines an equilateral triangle, i.e., $\triangle OAB$, $\triangle OBC$ and
$\triangle OCA$, seen in Fig.~\ref{N123} (c). These three triangles are the surfaces separating allowed and forbidden
regions. Therefore the inequalities have excluded the occupation of tetrahedra $ROAB$, $QOBC$, and $POCA$ from the entire cube. On the other hand, the inhabitable region resulting from the constraints by the three inequalities is simply the base-to-base union of the regular tetrahedron $OABC$ and the rectangular tetrahedron $MABC$. The entanglements $\{E_j\}$ of all physical quantum states have to be restricted to this allowed confined region. 

For all $N\ge4$ the restricted region defined by the restriction (\ref{polygon-ineq}) is a polytope, a hypervolume that is compact inside the unit
hypercube. In general, for any $N$, each individual inequality of (\ref{polygon-ineq}) excludes a rectangular simplex of the hypercube with a hypervolume given
by \cite{Cho}:
\begin{equation}
\prod_{j=1}^{N}\int_{0}^{1}[E_{j}]^{j-1}dE_{j}=\frac{1}{N!}.
\end{equation}
For example, in the three qubit case illustrated in Fig.~\ref{N123} (c), one of the excluded rectangular simplexes is tetrahedron $PAOC$ whose volume is simply $1/6$. Therefore the total available hypervolume, given the restrictions by all $N$ such inequalities, is
\begin{equation}\label{availablehypervolume}
V_N=1-\frac1{(N-1)!}.
\end{equation}
One can easily check for the three-qubit case that the volume being allowed is 1/2. 

According to Eq.~(\ref{availablehypervolume}), the ratio of the
allowed hypervolume $V_N$ to the unit hypervolume increases as the number
of qubits is increased, approaching unity as $N\to\infty$. That is, the more qubits that exist in the system, the less restriction there will be among all marginal entanglements, and the more sharing flexibility there will be (the issue of sharing will be addressed in the following).  This may be viewed as an advantage of using multiparty systems in the realization of quantum information tasks in the sense that all the entanglements existing in the system are more flexible than in fewer-party systems. 

A variation of the entanglement polygon inequality (\ref{polygon-ineq}) is the relation (\ref{sharing}), which reveals an important aspect of entanglement resource sharing. That is, it reveals a rule how to share a given amount of total entanglement $E_T$. 
%This rule (\ref{sharing}) can be easily interpreted as: no individual $E_j$ should get more than half of the total resource. 
In principle, after obeying this rule, there may still be some flexibility allowed for sharing a given amount of total resource. We now analyze quantitatively the effect of (\ref{sharing}) on sharing flexibility or sharing capacity.

\begin{figure}[h!]
\includegraphics[width=7.5cm]{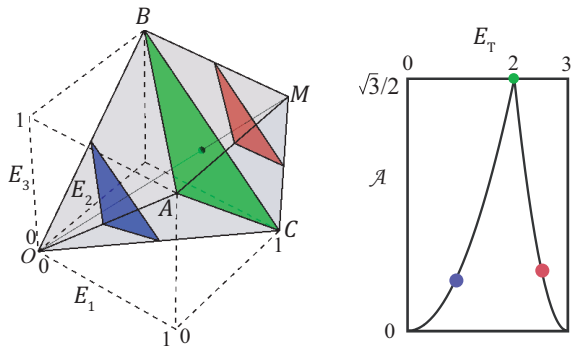}
\caption{Inhabitable region inside the unit entanglement cube. The entanglement polygon inequality (\ref{polygon-ineq}) confines the inhabitable region to just two tetrahedra $OABC$ and $MABC$ (shaded in gray). Also shown are three triangular planar sections of this region transverse to the unit cube's body diagonal. Degree of sharing ${\cal A}$ is shown as a function of $E_{\rm T}$. The three colored dots correspond respectively to the three
colored triangles in panel (a). }
\label{tetrahedron}
\end{figure}

The geometric representation helps to visualize the freedom of distributing entanglements. We start
by noticing in the $N=3$ case that the domains of different total entanglements $E_T$ define triangles transverse to
the body diagonal (color triangles in Fig.~\ref{tetrahedron}) under the inequality restriction.
Inspection shows that the $E_{T}$ value for these triangles varies from 0 to 3, running from zero to maximal total entanglement. It is obvious that infinitely many combinations of ${\bf E}=(E_1, E_2, E_2)$ are available to sum to the total $E_T$ in each transverse triangle, which makes it difficult to quantify.

An advantage of the geometric representation is that it allows us to adopt the area ${\cal A}$ of each triangle as a natural quantitative measure of entanglement-sharing capacity. The relation between ${\cal A}$ and the amount of total entanglement to be shared is not a linear relation, but a piece-wise quadratic of the form:
\beqa
{\cal A} &=& \frac{\sqrt{3}}{2} \times \left\{\begin{array}{cc}
E_{T}^{2}/4, & 0 \leq E_{T} \leq 2,\\
(3-E_{T})^{2}, & 2\leq E_{T}\leq 3.
\end{array}\right.
\eeqa
The sharing capacity ${\cal A}$ is graphed in Fig.~\ref{tetrahedron}, where we see that it is peaked around its maximum of $\sqrt{3}/2$ at $E_{\rm T}=2$, corresponding to the triangle $\triangle ABC$. It should be noted that greater total entanglement $E_T$ does not guarantee greater sharing capacity.

One can further define the sharing capacity for the $N$-qubit case as the hyperarea of the $(N-1)$-dimensional inhabitable polytope of fixed $E_{\rm T}$, normal to the line $OM$ within the $N$-dimensional polytope restricted by
the $N$ inequalities in the form of (\ref{sharing}). Again $O$ is the point of  zero total entanglement and $M$ represents maximum total entanglement. The general area expression is given as
\beqa
{\cal A} &=& \sqrt{N} \times \left\{\begin{array}{cc}
(1-\frac{N}{2^{N-1}})\frac{E_{\rm T}^{N-1}}{(N-1)!}, & E_{\rm T} \le 2,\\
{\cal B}_{N-1}(E_{\rm T}), & E_{\rm T}\ge 2,
\end{array}\right.
\eeqa
where ${\cal B}_{N-1}(E_{\rm T})$ is the zeroth uniform B-spline basis function \cite{Knott} of degree $N-1$ at $E_{\rm T}$ with knots $E_T=2,3,4,...,N$. It is known that these functions provide the diagonal cross-sections of $N$-dimensional hypercubes \cite{Warren-Weimer}. Note that inequality (2) only affects the interval $0\le E_{\rm T}\le2$, within which the corresponding result can be calculated directly. Here ${\cal A}$ is a piecewise polynomial of $E_{\rm T}$ of order $N-1$, which vanishes at the endpoints $E_{\rm T}=0$ (corresponding to point $O$) and $E_{\rm T}=N$ (corresponding to point $M$).

It is worth to note that the parameters of our geometric representation are different entanglements. This differs from the direct focus on quantum state parameters; see for example a recent geometric analysis of quantum state discrimination \cite{Bergou}. However, it would still be very interesting to explore the connection between quantum states and our entanglement representation; a brief discussion is now presented as an illustration. \\

%\section{example}
\noindent{\bf Example of Entangled States:} Let us now view our results with specific examples by considering the normalized Schmidt weight $Y_j$ for three-qubit generalized GHZ \cite{GHZ-07} class states and its inequivalent W \cite{DVC-00} class states:
\beq \label{GHZ}
|\Psi_{\rm GHZ}\rangle = \cos\theta |0,0,0\ra + \sin\theta |1,1,1\ra,
\eeq
and
\beq \label{W}
|\Psi_{\rm W}\rangle = \alpha |1,0,0\ra + \beta|0,1,0\ra + \gamma|0,0,1\ra.
\eeq

It is straightforward to note that the three marginal entanglements are given as $Y_1=Y_2=Y_3=1-|\cos2\theta|$ for the GHZ-class states. Obviously, they satisfy the polygon inequality (\ref{polygon-ineq}). In the geometric representation, these states live along the cube's body diagonal line $OM$ (see Fig.~\ref{tetrahedron}). One also sees that there is minimum sharing capacity ${\cal A}=0$ for the GHZ-class states simply because for any given total resource $Y_T$ there is only one way to share, i.e., $Y_1=Y_2=Y_3=Y_T/3$.

To investigate W-class states one needs to analyze all combinations of $|\alpha|, |\beta|, |\gamma|$, and consider four different cases, i.e., $|\alpha|^2\ge 1/2$, $|\beta|^2\ge 1/2$, $|\gamma|^2\ge 1/2$, and ${\rm Max} (|\alpha|^2,\ |\beta|^2,\  |\gamma|^2) < 1/2$. When $|\alpha|^2\ge 1/2$, one can compute the entanglements respectively as $Y_1=2(|\beta|^2+|\gamma|^2)$, $Y_2=2|\beta|^2$, and $Y_3=2|\gamma|^2$. This clearly satisfies the polygon inequality (\ref{polygon-ineq}). Particularly, this case satisfy one of the boundary equalities, $Y_1=Y_2+Y_3$, corresponding to $\triangle OBC$ in the geometric representation illustrated in Fig.~\ref{tetrahedron}. Similarly, the symmetric cases $|\beta|^2\ge1/2$ and $|\gamma|^2\ge1/2$ satisfy respectively the remaining two equalities $Y_2=Y_1+Y_3$ and $Y_3=Y_1+Y_2$. These correspond to $\triangle OCA$ and $\triangle OAB$ in the cube (see Fig.~\ref{tetrahedron}). The occupation of these boundaries indicates the unique tightness of our inequalities (\ref{polygon-ineq}) in terms of $Y$. 

When ${\rm Max} (|\alpha|^2,\ |\beta|^2,\  |\gamma|^2) \le1/2$, one has $Y_1=2|\alpha|^2$, $Y_2=2|\beta|^2$, $Y_3=2|\gamma|^2$, satisfying relation (\ref{polygon-ineq}). More interestingly, one sees $Y_1+Y_2+Y_3=2$. Therefore, in this case, the W-class states occupy the entire triangle  $\triangle ABC$ in the cube, which exhibits the maximum sharing capacity with ${\cal A}=\sqrt{3}/2$. \\

%############################## %\section{Summary}
%\section{Summary}
\noindent{\bf Summary:} We have presented an entanglement polygon inequality for arbitrary $N$-qubit pure states, and analyzed its restriction and sharing properties with a geometric representation. Its intimate connection to the well-known monogamy relation is also shown. 

The entanglement polygon inequality reveals a type of fundamental constraint among multiple entanglements of a multiparty system governed by quantum mechanics. It further establishes a resource sharing rule that limits the flexibility of distributing entanglements among all participants. Such a sharing rule may provide guidance when proposing optimal schemes that can run multiple entanglement-assisted quantum tasks in a single multiparty system.

Our geometric representation of the inequalities provides a potentially useful way to analyze and understand multiple entanglement restrictions, as well as to study collective or dynamical entanglement behavior. It has the advantage of exposing quantitatively the degree of restriction and sharing capacity. An example of its utility has already been demonstrated in the development of a center-of-mass interpretation of bipartite purity for both pure and mixed states (see \cite{AQE-16}).

Preliminary numerical results support the speculation that the same inequalities of $Y_j$ hold for pure states of multi-party $M$-level systems, where the normalized entanglement monotone becomes $Y_{j} = 1 -\sqrt{\frac{M-K_{j}}{K_{j}(M-1)}}$ and $K_j$ is the Schmidt weight of the extended pure state. This will simply extend our resource sharing treatment to a much wider category of quantum states. \\

%##############################

%\section{Acknowledgements} 
\noindent{\bf Acknowledgements:} We acknowledge partial financial
support from the National Science Foundation through awards PHY-1068325, PHY-1203931, PHY-1505189, PHY-1507278, and INSPIRE PHY-1539859. MAA received funding from the Excellence Initiative of Aix-Marseille University - A$^*$MIDEX, a French ``Investissements d'Avenir'' programme. XFQ would also like to thank Michael Hall and Shuming Chen for pointing out the connections between our entanglement constraint relations and the polygon inequalities.

%\noindent{\bf Proof of Entanglement Restriction:} 

\section{Appendix: Proof of Entanglement Polygon Inequality} 

To prove the entanglement polygon inequality, we first prove that the normalized Schmidt weight $Y$ \cite{Qian-Eberly-10} satisfies the set of inequalities in (\ref{polygon-ineq}). We then show that all other measures, i.e., von Neumann entropy \cite{von-Neumann-32}, concurrence \cite{Wootters-98}, and negativity \cite{Vidal-Werner-02} satisfy automatically the same symmetric inequality relation due to the fact that they are all concave and monotonically increasing functions of $Y$. \\

\noindent{\bf Polygon Inequality for $Y$:} First, we prove the entanglement inequalities in relation (\ref{polygon-ineq}) of the main text in terms of $Y$ for arbitrary $N$-qubit pure states. When bipartitioned between a single qubit (e.g., the $i$-th qubit) and remaining $N-1$ qubits, an $N$-qubit pure state can always be decomposed into the Schmidt form, i.e.,
\begin{equation}
|\Psi \rangle =\sqrt{\lambda _{1}^{(i)}}|f_{1}^{(i)}\rangle \otimes
|g_{1}^{(i)}\rangle +\sqrt{\lambda _{2}^{(i)}}|f_{2}^{(i)}\rangle \otimes
|g_{2}^{(i)}\rangle,  \label{Schmidt decompose}
\end{equation}
where $|f_{n}^{(i)}\rangle $\ and $|g_{n}^{(i)}\rangle$, $n=1,2$, are the
Schmidt bases of $i$-th qubit and the remaining $N-1$ qubits respectively. Here $\lambda _{1}^{(i)}$ and $\lambda _{2}^{(i)}$ are the
corresponding Schmidt coefficients and we assume $\lambda _{1}^{(i)}\geq \lambda _{2}^{(i)}$ for all $i$
without loss of generality.

For simplicity we prove only the first inequality:
\begin{equation}
\sum_{j=2}^{N}Y_{j}\geq Y_{1};  \label{inequality1}
\end{equation}
the remaining $N-1$ inequalities follow by symmetry.
We hence consider the specific Schmidt decomposition
with respect to qubit 1 by taking $i=1$ in (\ref{Schmidt decompose}). We express the two $(N-1)$-qubit states $|g_{1}^{(1)}\rangle$ and $|g_{2}^{(1)}\rangle$ in the Schmidt basis of each
qubit, with complex amplitudes $x_{j}$ and  $y_{j}$, i.e.,
\begin{subequations}\label{grels}
\begin{eqnarray}
|g_{1}^{(1)}\rangle &=&x_{1}|f_{1}^{(2)}\rangle ...|f_{1}^{(i)}\rangle
...|f_{1}^{(N)}\rangle  \notag \\
&&+x_{2}|f_{1}^{(2)}\rangle ...|f_{1}^{(i)}\rangle ...|f_{2}^{(N)}\rangle
\notag \\
&&+...+x_{2^{(N-1)}}|f_{2}^{(2)}\rangle ...|f_{2}^{(i)}\rangle
...|f_{2}^{(N)}\rangle,  \label{g11}
\\
|g_{2}^{(1)}\rangle &=&y_{1}|f_{1}^{(2)}\rangle ...|f_{1}^{(i)}\rangle
...|f_{1}^{(N)}\rangle  \notag \\
&&+y_{2}|f_{1}^{(2)}\rangle ...|f_{1}^{(i)}\rangle ...|f_{2}^{(N)}\rangle
\notag \\
&&+...+y_{2^{(N-1)}}|f_{2}^{(2)}\rangle ...|f_{2}^{(i)}\rangle
...|f_{2}^{(N)}\rangle,  \label{g12}
\end{eqnarray}
\end{subequations}
where we have the orthonormality conditions
\begin{subequations}
\begin{eqnarray}
\sum_{j=1}^{2^{N-1}}|x_{j}|^{2} &=&\sum_{j=1}^{2^{N-1}}|y_{j}|^{2}=1,
\label{condition1} \\
\sum_{j=1}^{2^{N-1}}x_{j}y_{j}^{\ast } &=&0.  \label{condition2}
\end{eqnarray}
\end{subequations}
The entanglement between qubit 1 and the remaining qubits can then be
easily obtained as
\begin{equation}
Y_{1}=2\lambda _{2}^{(1)}.  \label{Y1}
\end{equation}

Now we rearrange the state (\ref{Schmidt decompose}), and write it by grouping the states of
qubit 2, i.e.,
\begin{eqnarray}
|\Psi \rangle &=&|f_{1}^{(2)}\rangle \Big[ \sqrt{\lambda _{1}^{(1)}}x_{1}|f_{1}^{(1)}\rangle |f_{1}^{(3)}\rangle ...|f_{1}^{(N)}\rangle
\notag \\
&&+...+\sqrt{\lambda _{1}^{(1)}}x_{2^{(N-2)}}|f_{1}^{(1)}\rangle |f_{2}^{(3)}\rangle
...|f_{2}^{(N)}\rangle  \notag \\
&&+\sqrt{\lambda _{2}^{(1)}}y_{1}|f_{2}^{(1)}\rangle |f_{1}^{(3)}\rangle
...|f_{1}^{(N)}\rangle  \notag \\
&&+...+\sqrt{\lambda _{2}^{(1)}}y_{2^{(N-2)}}|f_{2}^{(1)}\rangle |f_{2}^{(3)}\rangle
...|f_{2}^{(N)}\rangle \Big]  \notag \\
&&+|f_{2}^{(2)}\rangle \Big[ \sqrt{\lambda _{1}^{(1)}}x_{2^{(N-2)}+1}|f_{1}^{(1)}\rangle
|f_{1}^{(3)}\rangle ...|f_{1}^{(N)}\rangle  \notag \\
&&+...+\sqrt{\lambda _{1}^{(1)}}x_{2^{(N-1)}}|f_{1}^{(1)}\rangle |f_{2}^{(3)}\rangle
...|f_{2}^{(N)}\rangle  \notag \\
&&+\sqrt{\lambda _{2}^{(1)}}y_{2^{(N-2)}+1}|f_{2}^{(1)}\rangle |f_{1}^{(3)}\rangle
...|f_{1}^{(N)}\rangle  \notag \\
&&+...+\sqrt{\lambda _{2}^{(1)}}y_{2^{(N-1)}}|f_{2}^{(1)}\rangle |f_{2}^{(3)}\rangle
...|f_{2}^{(N)}\rangle \Big].  \label{Schmidt2}
\end{eqnarray}
It is easy to note that the corresponding Schmidt coefficients for qubit 2
are given as
\begin{eqnarray}
\lambda _{1}^{(2)} &=&\sum_{j=1}^{2^{(N-2)}}(\lambda
_{1}^{(1)}|x_{j}|^{2}+\lambda _{2}^{(1)}|y_{j}|^{2}), \\
\lambda _{2}^{(2)} &=&\sum_{j=1}^{2^{(N-2)}}(\lambda
_{1}^{(1)}|x_{2^{(N-2)}+j}|^{2}+\lambda _{2}^{(1)}|y_{2^{(N-2)}+j}|^{2}).
\end{eqnarray}
Again the entanglement measure between qubit 2 and the rest is obtained as
\begin{equation}
Y_{2}=2\sum_{j=1}^{2^{(N-2)}}(\lambda
_{1}^{(1)}|x_{2^{(N-2)}+j}|^{2}+\lambda _{2}^{(1)}|y_{2^{(N-2)}+j}|^{2}).
\label{Y2}
\end{equation}
We note that the expression for $Y_{2}$ simply
picks up the coefficients $|x_{j}|^{2}$ and $|y_{j}|^{2}$ that correspond to
the $(N-1)$-qubit basis states, as given in Eqs.~(\ref{grels}), when qubit 2 is in the Schmidt basis $|f_{2}^{(2)}\rangle $.
Similarly, the $i$-th qubit $Y_{i}$, which can be expressed in similar form as (\ref{Y2}), picks all the coefficients $|x_{j}|^{2}$ and $|y_{j}|^{2}$ that correspond to the $(N-1)$-qubit basis states containing $|f_{2}^{(i)}\rangle $. When summing over all $Y_{i}$ from $2$ to $N$, we get
\begin{eqnarray}
\sum_{j=2}^{N}Y_{j} &=&2\lambda
_{1}^{(1)}\Big[|x_{2}|^{2}+|x_{3}|^{2}+...+|x_{N}|^{2}  \notag \\
&&+2|x_{N+1}|^{2}+...+(N-1)x_{2^{(N-1)}}\Big]  \notag \\
&&+2\lambda _{2}^{(1)}\Big[|y_{2}|^{2}+|y_{3}|^{2}+...+|y_{N}|^{2}  \notag \\
&&+2|y_{N+1}|^{2}+...+(N-1)y_{2^{(N-1)}}\Big].  \label{Ysum}
\end{eqnarray}
That is, the number of times $|x_{j}|^{2}$ appears in the sum equals the number of times $|f_{2}^{(i)}\rangle$ appears in the corresponding $(N-1)$-qubit basis states given in Eqs.~(\ref{grels}).

From the above summation, along with the assumption $\lambda _{1}^{(1)}\geq
\lambda _{2}^{(1)}$, one immediately finds
\begin{eqnarray}
\sum_{j=2}^{N}Y_{j} &\geq &\sum_{k=2}^{2^{N-1}}2(\lambda
_{1}^{(1)}|x_{k}|^{2}+\lambda _{2}^{(1)}|y_{k}|^{2}),  \notag \\
&\geq &\sum_{k=2}^{2^{N-1}}2\lambda _{2}^{(1)}(|x_{k}|^{2}+|y_{k}|^{2}).
\label{simplified1}
\end{eqnarray}
We note that in order for Eq.~(\ref{simplified1}) to hold, it needs each $|x_k|^2$ and $|y_k|^2$ inside the square brackets of Eq.~(\ref{Ysum}) to have a coefficient greater than or equal to 1. If the assumption $\lambda _{1}^{(i)}\geq \lambda_{2}^{(i)}$ is removed, one will get a different version of Eq. (\ref{Ysum}) where the actual coefficients for each $|x_k|^2$ and $|y_k|^2$ inside the square brackets will be different from the current Eq.~(\ref{Ysum}). However, these coefficients are still determined by the number of times either $f_1$s or $f_2$s appear in Eq.~(\ref{grels}) in each particular vector, which guarantees that there are at least one $|x_k|^2$ and one $|y_k|^2$ for all $2\le k\le 2^{(N-1)}$ in Eq.~(\ref{Ysum}). This suffices to get Eq.~(\ref{simplified1}).

From relation (\ref{simplified1}) along with (\ref{Y1}), proving relation (\ref{inequality1}) requires only proving the following relation
\begin{equation}
1\geq |x_{1}|^{2}+|y_{1}|^{2}.  \label{simplified-inequality}
\end{equation}
From condition (\ref{condition2}), one has
\begin{equation}
|x_{1}y_{1}^{\ast }|^{2}=\left|x_{2}y_{2}^{\ast
}+\sum_{j=3}^{2^{(N-1)}}x_{j}y_{j}^{\ast }\right|^{2}.  \label{orthogonal relation}
\end{equation}
The right hand side (RHS) of (\ref{orthogonal relation}) can be written as
\begin{eqnarray}
{\rm RHS} &=&\left(x_{2}y_{2}^{\ast }+\sum_{j=3}^{2^{(N-1)}}x_{j}y_{j}^{\ast
}\right)\left(x_{2}^{\ast }y_{2}+\sum_{j=3}^{2^{(N-1)}}x_{j}^{\ast }y_{j}\right)  \notag \\
&=&|x_{2}y_{2}|^{2}+x_{2}y_{2}^{\ast }\sum_{j=3}^{2^{(N-1)}}x_{j}^{\ast }y_{j}
\notag \\
&&+x_{2}^{\ast }y_{2}\sum_{j=3}^{2^{(N-1)}}x_{j}y_{j}^{\ast
}+\sum_{j=3}^{2^{(N-1)}}x_{j}y_{j}^{\ast }\sum_{j=3}^{2^{(N-1)}}x_{j}^{\ast
}y_{j}.
\end{eqnarray}
By using condition (\ref{condition1}), one can write this expression as
\begin{eqnarray}
{\rm RHS}
&=&\left(1-|x_{1}|^{2}-\sum_{j=3}^{2^{(N-1)}}|x_{j}|^{2}\right)\left(1-|y_{1}|^{2}-
\sum_{j=3}^{2^{(N-1)}}|y_{j}|^{2}\right)  \notag \\
&&+x_{2}y_{2}^{\ast }\sum_{j=3}^{2^{(N-1)}}x_{j}^{\ast }y_{j}+x_{2}^{\ast
}y_{2}\sum_{j=3}^{2^{(N-1)}}x_{j}y_{j}^{\ast }  \notag \\
&&+\sum_{j=3}^{2^{(N-1)}}x_{j}y_{j}^{\ast }\sum_{j=3}^{2^{(N-1)}}x_{j}^{\ast
}y_{j}, \notag\\
&=&1-|y_{1}|^{2}-\sum_{j=3}^{2^{(N-1)}}|y_{j}|^{2}-|x_{1}|^{2}+|x_{1}y_{1}|^{2}
\notag \\
&&+|x_{1}|^{2}\sum_{j=3}^{2^{(N-1)}}|y_{j}|^{2}-
\sum_{j=3}^{2^{(N-1)}}|x_{j}|^{2}+|y_{1}|^{2}\sum_{j=3}^{2^{(N-1)}}|x_{j}|^{2}
\notag \\
&&+\sum_{j=3}^{2^{(N-1)}}|x_{j}|^{2}%
\sum_{j=3}^{2^{(N-1)}}|y_{j}|^{2}+x_{2}y_{2}^{\ast
}\sum_{j=3}^{2^{(N-1)}}x_{j}^{\ast }y_{j}  \notag \\
&&+x_{2}^{\ast }y_{2}\sum_{j=3}^{2^{(N-1)}}x_{j}y_{j}^{\ast
}+\sum_{j=3}^{2^{(N-1)}}x_{j}y_{j}^{\ast }\sum_{j=3}^{2^{(N-1)}}x_{j}^{\ast
}y_{j}.
\end{eqnarray}
When comparing with the left hand side of (\ref{orthogonal relation}), one
immediately has
\begin{equation}
1-|x_{1}|^{2}-|y_{1}|^{2}=\Delta,\label{Delta}
\end{equation}
where, following some algebra, $\Delta$ can expressed as
\begin{eqnarray}
\Delta
&=&|x_{2}|^{2}\sum_{j=3}^{2^{(N-1)}}|y_{j}|^{2}+(|y_{2}|^{2}+
\sum_{j=3}^{2^{(N-1)}}|y_{j}|^{2})\sum_{j=3}^{2^{(N-1)}}|x_{j}|^{2}  \notag \\
&&-x_{2}y_{2}^{\ast }\sum_{j=3}^{2^{(N-1)}}x_{j}^{\ast }y_{j}-x_{2}^{\ast
}y_{2}\sum_{j=3}^{2^{(N-1)}}x_{j}y_{j}^{\ast }  \notag \\
&&-\sum_{j=3}^{2^{(N-1)}}x_{j}y_{j}^{\ast }\sum_{j=3}^{2^{(N-1)}}x_{j}^{\ast
}y_{j},
\nonumber\\
&=&|x_{2}|^{2}\sum_{j=3}^{2^{(N-1)}}|y_{j}|^{2}+|y_{2}|^{2}
\sum_{j=3}^{2^{(N-1)}}|x_{j}|^{2}  \notag \\
&&-x_{2}y_{2}^{\ast }\sum_{j=3}^{2^{(N-1)}}x_{j}^{\ast }y_{j}-x_{2}^{\ast
}y_{2}\sum_{j=3}^{2^{(N-1)}}x_{j}y_{j}^{\ast }  \notag \\
&&+\sum_{j=3}^{2^{(N-1)}}|y_{j}|^{2}\sum_{j=3}^{2^{(N-1)}}|x_{j}|^{2}-
\sum_{j=3}^{2^{(N-1)}}x_{j}y_{j}^{\ast }\sum_{j=3}^{2^{(N-1)}}x_{j}^{\ast }y_{j},
\notag \\
&=&\sum_{j=3}^{2^{(N-1)}}|x_{2}y_{j}-y_{2}^{\ast }x_{j}^{\ast
}|^{2}+\Big(\sum_{j>k\ge3}^{2^{(N-1)}}+\sum_{k>j\ge3}^{2^{(N-1)}}+\sum_{j=k\ge3}^{2^{(N-1)}}\Big)\notag \\
&& (|x_{j}|^{2}|y_{k}|^{2}-x_{j}y_{j}^{\ast }x_{k}^{\ast }y_{k}) \notag \\
&=&\sum_{j=3}^{2^{(N-1)}}|x_{2}y_{j}-y_{2}^{\ast }x_{j}^{\ast
}|^{2}+\sum_{j>k\ge3}^{2^{(N-1)}}|x_{j}y_{k}-x_{k}^{\ast }y_{j}^*|^{2},
\end{eqnarray}
where we used the fact that $\sum_{j=k\ge3}^{2^{(N-1)}}(|x_{j}|^{2}|y_{k}|^{2}-x_{j}y_{j}^{\ast }x_{k}^{\ast }y_{k})=0$. Obviously, $\Delta\ge0$, which indicates that Eq.~(\ref{Delta}) provides the desired proof of relation (\ref{simplified-inequality}) and consequently the inequality (\ref{inequality1}). The proof will be exactly symmetric for all other inequalities in equation (\ref{polygon-ineq}) of the main text. In these cases one will just have to prove an inequality similar to (\ref{simplified-inequality}), but by replacing $|x_{1}|^{2}$, $|y_{1}|^{2}$ with $|x_{i}|^{2}$, $|y_{i}|^{2}$.\\

%##############################

\noindent{\bf Polygon Inequality for other measures:} From the definitions of the the entanglement measures von Neumann entropy $S$, Concurrence $C$, and normalized Negativity $N$, it is straightforward to express them in terms of  the normalized Schmidt weight $Y$, i.e., 
\beqa \label{efield}
S(Y)&=&1-[(2-Y)\log_2 (2-Y)+Y\log_2Y]/2,\notag\\
C(Y)&=& \sqrt{Y(2-Y)},\notag \label{C-Y}\\
N(Y)&=& \sqrt{Y(2-Y)}.\notag
\eeqa
One sees that they are all monotonic increasing concave functions of $Y$ in the region [0,1].

%\beqa \label{efield}
%{\rm von\ Neumann\ } S&=&1-(Y\log_2Y)/2  \notag \\ 
%&-&(1-Y/2)\log_2 (2-Y),\\
%{\rm Concurrence\ }  C&=& \sqrt{Y(2-Y)},\\
%{\rm Negativity\ } N&=& \sqrt{Y(2-Y)},\\
%{\rm Schmidt Weight\ } K&=&2/[1+(1-Y)^2].
%\eeqa

We first assume that Max$\{Y_i\}=Y_j$, with $i=1,2,3,...,N$. Then one immediately has Max$\{E(Y_i)\}=E(Y_j)$ due the monotonic increasing property of the function $E(Y)$. Obviously, this will lead to the following relation 
\beq 
E(Y_i)\le E(Y_1)+...+E(Y_{k\neq i})+...+E(Y_N),
\eeq
for any $i\ne j$. Therefore, what needs to be proved is only one relation, i.e.,
\beq \label{target}
E(Y_j)\le E(Y_1)+...+E(Y_{k\neq j})+...+E(Y_N).
\eeq

\begin{figure}[h]
\includegraphics[width=5.5cm]{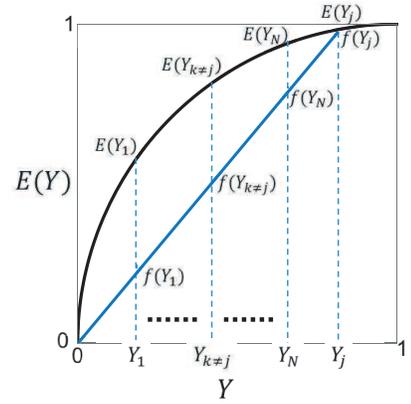}
\caption{Schematic illustration of a concave and monotonically increasing function $E(Y)$, as well as a linear function $f(Y)=E(Y_j)Y/Y_j$ of the parameter $Y$. In the plot $E(Y)$ is taken as $\sqrt{Y(2-Y)}$ which represents concurrence and negativity.}
\label{E(Y)}
\end{figure}

To prove the above relation we use a geometric illustration, as shown in Fig.~\ref{E(Y)}, for visualization assitance. The black solid line is a generic $E(Y)$ function which is concave and monotonically increasing with respect to $Y$. The blue solid line is a linear function of $Y$ defined as 
\beq
f(Y)=\frac{E(Y_j)}{Y_j}Y.
\eeq
It crosses with $E(Y)$ at $Y=Y_j$, $f(Y_j)=E(Y_j)$ and $Y=0$, $f(0)=E(0)=0$.

First, we consider the sum of all $f(Y)$ values with respect to $N$ entanglement $Y$ values determined by the $N$-qubit system. That is,
\beqa
\sum _k f(Y_{k\ne j})=\frac{\sum_k Y_{k\ne j}}{Y_j}E(Y_j)\ge E(Y_j),
\eeqa
where we have used the fact that $Y_j\le \sum_k Y_{k\ne j}$.

Second, from the concavity of $E(Y)$, one immediately sees that $E(i)\ge f(Y_{i})$ for any $i=1,2,...,N$, as illustrated in Fig.~\ref{E(Y)}. This leads directly to the relation
\beqa
\sum_k E(Y_{k\ne j})\ge \sum_k f(Y_{k\ne j})\ge E(Y_j),
\eeqa
which is exactly (\ref{target}). To this end we have proved that any concave and monotonically increasing function $E(Y)$ with respect to $Y\in[0,1]$ will satisfy a similar polygon inequality in terms of (\ref{target}) for any $Y_j$.

In the literature, there exist many other entanglement measures beside the von Neumann entropy $S$, concurrence $C$, and negativity $N$; see for example an overview in Ref.~\cite{Plenio-Virmani}. It would be interesting to check whether other measures will also satisfy the same polygon inequality (\ref{polygon-ineq}). We expect that several of them are also concave and monotonically increasing functions of $Y$ so that they satisfy the same relation (\ref{polygon-ineq}) immediately. \\

\end{document}